\documentclass[12pt]{iopart}
\usepackage{iopams}  
\usepackage{setstack}

\begin{document}
%\title[Loop groups and integrable quad systems]{Loop groups and integrable quad systems}
\title[An algebraic criterion of the Darboux integrability ]{An algebraic criterion of the Darboux integrability of differential-difference equations and systems}

\author{I T Habibullin, M N Kuznetsova}

\address{Institute of Mathematics, Ufa Federal Research Centre, Russian Academy of Sciences,
112 Chernyshevsky Street, Ufa 450008, Russian Federation}
\eads{\mailto{habibullinismagil@gmail.com}, \mailto{mariya.n.kuznetsova@gmail.com}}

\begin{abstract}
The article investigates systems of differential-difference equations of hyperbolic type, integrable in sense of Darboux. The concept of a complete set of independent characteristic integrals underlying Darboux integrability is discussed. A close connection is found between integrals and characteristic Lie-Rinehart algebras of the system. It is proved that a system of equations is Darboux integrable if and only if its characteristic algebras in both directions are finite-dimensional.
\end{abstract}

\submitto{J. Phys. A: Math. Theor.}
\maketitle

\eqnobysec

\section{Introduction}

The article deals with systems of differential-difference equations of the following form:
\begin{equation}	\label{main_sys}
u^j_{n+1,x} = F^j(x,n,u^j_{n,x}, u_n, u_{n+1}), \quad j=1,2, \ldots, N,
\end{equation}
where $u_n = (u^1_n, u^2_n, \ldots, u^N_n)$ is the sought function, its components  $u^j_n = u^j_n(x)$ depend on two variables: real $x$ and integer $n$. It is assumed that $F^j$ analytically depends on the dynamical variables $u^j_{n,x}$, $u_n$, $u_{n+1}$ in some domain of the space $\mathbb{C}^{2N+1}$. We study the system from the point of view of Darboux integrability. Below we will briefly explain the essence of this concept. Fundamental ideas in the study of the problem of constructing in closed form general solution of partial differential equations of hyperbolic type go back to the classical works of Laplace, Liouville, Lie, Darboux, Goursat, Vessio, and others. One of the basic approaches to the problem is worked out by Darboux. The first step in Darboux's method is to find integrals over each characteristic direction, which gives two ordinary differential equations. In the second one a formula is looked for parameterizing general solutions of these ODEs. Let us explain Darboux's method using the example of the well known Liouville equation $u_{xy}=e^u$ (see, for instance, \cite{ZhS2001}). It has integrals of the form 
$$I=u_{xx}-\frac{1}{2}u_x^2, \qquad  J=u_{yy}-\frac{1}{2}u_y^2.$$
Functions $I$ and $J$ are such that for arbitrary solution $u(x,y)$ of the Liouville equation the equalities are fulfilled: $\frac{d}{dy}I=0$ and $\frac{d}{dx}J=0$. That immediately implies that each solution of the equation satisfies a pair of the ordinary differential equations 
\begin{equation} \label{ode}
u_{xx}-\frac{1}{2}u_x^2=p(x), \qquad  u_{yy}-\frac{1}{2}u_y^2=q(y).
\end{equation}
It is noteworthy that the representation (\ref{ode}) of an arbitrary solution makes it possible to derive a well-known formula that parametrizes the general solution of the Liouville equation
\begin{equation} \label{solution}
u(x,y)=\log \left(\frac{2\phi'(x)\psi'(y)}{(\phi(x)+\psi(y))^2}\right) .
\end{equation}

It is not surprising that Liouville-type equations admit an explicit formula for the general solution. Actually existence of a nontrivial integral makes it possible to find a differential substitution that reduces the equation to the d'Alembert equation. In the case if the Liouville equation it is done easily. Indeed, by taking $v_x=u_{xx}-\frac{1}{2}u_x^2$ we obtain that $v(x,y)$ due to formula (\ref{ode}) satisfies equation $v_{xy}=0$.

To find the integrals (as well as to find out the integrability of a given equation) Darboux used Laplace cascade method. In later studies (see \cite{Goursat1}, \cite{Goursat2}, \cite{ves1}, \cite{ves2}, the direct method using characteristic vector fields becomes the main tool for searching for integrals (it is within this approach that
apparently, the first lists of equations with integrals in both directions  were obtained, see \cite{Goursat1}, \cite{Goursat2}). 

Interest in this topic revived at the end of the twentieth century after the works \cite{ShabatYam81, LSS}, where the concept of characteristic algebra was introduced, which became an effective implement for studying and classifying systems of partial differential equations of hyperbolic type. The fundamental fact of this theory is that a system of PDEs is Darboux integrable if and only if its characteristic algebras in both characteristic directions are of finite dimension (see \cite{ZMHS}). Note that characteristic algebras of the exponential type systems of differential equations are actually Lie algebras over the field of constants. But later it was observed that in general characteristic algebra is really an algebra over the ring of locally analytic functions, such kind generalization  of the characteristic algebra was used for the sake of the integrable classification of differential-difference equations in \cite{HPZ2009}. We thank D.Millionschikov who pointed out that such kind  algebra is a well known object called in literature a Lie-Rinehart algebra \cite{Rinehart, Million, MillionSmirnov}.

At present, there are various alternative approaches to the problem of classifying equations of the Liouville type, such as the method of higher symmetries, the method of Laplace invariants, etc. Interesting classification results, a description of the current state of research in this area and references can be found in \cite{ZhS2001}, \cite{Anderson}.

Note that interest in discrete equations is steadily growing due to their important applications in physics, informatics, biology, etc. Various approaches to constructing solutions of such equations are often discussed in the literature. One of the purposes of this work is to transfer the main stages of the Darboux method to differential-difference systems of the form (\ref{main_sys}). It should be noted that different aspects of the Darboux integrability of differential-difference equations nowadays are actively discussed (see, for instance, \cite{AdlerStartsev1999}, \cite{Smirnov2015}, \cite{GarYam19}, \cite{GarHab21}, \cite{Demskoy10}, \cite{ZhZh21}, \cite{HPZ2009}). 

The second purpose of the article is related to a recently found application of the system (\ref{main_sys}) in the problem of the integrable classification of the lattices in 3D. In a paper \cite{HabibullinKhakimova} by A. Khakimova and one of the authors of this article, a method was proposed for the integrable classification of nonlinear equations in 3D 
\begin{equation} \label{eq0}  
u_{n+1,x}^j = F(u_{n,x}^j,u_n^{j+1},u_{n+1}^j,u_n^j,u_{n+1}^{j-1}),
\end{equation}
with two discrete and one continuous independent variables by reducing the problem to studying higher order differential-difference Darboux integrable systems with two independent variables. The above-mentioned method of integrable reductions is also an effective tool for constructing particular solutions of three-dimensional models of the form (\ref{eq0}). However, some important questions remained unexplored in \cite{HabibullinKhakimova}, such as the description of the structure of integrals of Darboux integrable differential-difference systems, as well as the proof of the algebraic criterion for Darboux integrability, although the criterion was formulated there as a conjecture. These gaps are filled in this work. 

In \cite{HabibullinKhakimova} it was observed that all known integrable equations with one continuous and two discrete independent variables, presented in \cite{FNR} can be written in the form  (\ref{eq0}), it was checked that these equations admit reductions being finite field Darboux integrable systems of semi-discrete hyperbolic type equations (\ref{main_sys}). Class of equations (\ref{eq0}) defines differential-difference versions of the famous Hirota type difference equations and is studied by many authors, the references can be found in \cite{FNR}.

A classification method based on the Darboux integrable reductions has earlier been used in our articles \cite{H2013,HP2017,HabKuzTMPh} for studying differential-difference equations of the form 
\begin{equation}  \label{Toda}  
u_{n,xy} = f(u_{n+1},u_n,u_{n-1}, u_{n,x},u_{n,y}),
\end{equation}
having one discrete and two continuous independent variable $n$, $x$ and $y$. Classification results concerned to the equations (\ref{Toda}) can be found in \cite{HabKuzTMPh, HKS2020, FHKN2020, HP2017}.

It worth mentioning that the integrability theory for the Liouville type systems of hyperbolic PDE has been developed earlier by many authors.  In the works  \cite{ZMHS}, \cite{KostriginaDiss}  such important aspects of the theory, as searching of the complete set  of integrals by using characteristic algebras,   integrability criterion, finding general solutions has been studied in detail. Below we discuss the similar problems for systems of the differential-difference equations. Some properties of these two objects are very resembling, in those cases we use the technique worked out in \cite{ZMHS}, \cite{KostriginaDiss}. However, there are aspects where new ideas are needed. For instance, the structure of the characteristic algebra for the second case is essentially more complex. 

The main result of this paper is the proof of the statement that for the Darboux integrability of the system of differential-difference equations (\ref{main_sys}) it is necessary and sufficient that the characteristic Lie-Rinehart algebras corresponding to both characteristic directions be of  finite dimension.

Let's briefly comment on the content of the article. In \S2 we give definitions of integrals of the system (\ref{main_sys}) for both characteristic directions $x$ and $n$, explain the essence of Darboux integrability. To describe the structure of the variety of integrals, we introduce the concept of a set of independent integrals of minimal orders. In fact, the set thus defined constitutes the basis of the integrals of the system
 (\ref{main_sys}). In \S3, we recall the definition of the characteristic Lie-Rinehart algebras in directions of $x$ and $n$, discuss some of their properties, and derive an algebraic criterion of existence of a complete set of $n$-integrals (see Theorems 3.2 and 3.3 ). The proof of Theorem 3.3 contains a detailed description of the method for constructing the set of integrals of minimal orders. Illustrative examples are given in \S 4.

\section{Basis of integrals}

In this section we discuss the problem of constructing bases of the integrals for the main object of study of the present article, the system of differential-difference hyperbolic type equations (\ref{main_sys}). 

Let us recall the necessary definitions and explain the basic notations. Variables $u^j_n, u^j_{n \pm 1}, u^j_{n \pm 2}, \ldots$ and $u^j_{n,x}, u^j_{n,xx}, u^j_{n,xxx}, \ldots$ are called dynamical variables and we consider them independent. We denote by $D_x$ the operator of the total derivative with respect to $x$; $D_n$ stands for the operator shifting the discrete argument $n$ such that $D_n y(n) = y(n+1)$. For higher-order derivatives with respect to $ x $, we use the notation:
\begin{equation*}
u^j_{n,[m]} = \frac{\partial^m u^j_n}{\partial x^m}.
\end{equation*}

{\bf Definition 2.1.}
{\it A function 
\begin{equation*}
I = I(x,n,u_n,u_{n,x}, u_{n,xx}, \ldots, u_{n, [m]})
\end{equation*}
is called an $n$-integral of the order $m$, if at least one of the derivatives $\frac{\partial I}{\partial u^j_{n,[m]}}$ where $j=1,2,\ldots,N$ does not vanish identically and the condition $D_n I = I$ is satisfied by virtue of system (\ref{main_sys}).}

Set of $n$-integrals $I^1, \ldots, I^N$ of the orders $n_1, n_2, \ldots, n_N$ is called independent if the set of the functions $D^j_x I^i$, where $i = 1, \ldots, N$, $j = 0,1, \ldots, s-n_i$ and $s = max \{n_i\}$ is functionally independent.

{\bf Definition 2.2.}
{\it A function
\begin{equation*}
J = J(x,n,u_{n-k}, u_{n-k+1}, \ldots, u_{n-1}, u_n, u_{n+1}, \ldots, u_{n+m-1}, u_{n+m} ), 
\end{equation*}
$k,m = 0,1,2,\ldots$ is called an $x$-integral of the order $m+k$, if at least for one pair of the numbers $j,s = 1, \ldots, N$ the product $\frac{\partial J}{\partial u^j_{n+m}} \dot \frac{\partial J}{\partial u^s_{n-k}}$ does not vanish identically and the following condition $D_x J = 0$ holds by virtue of system (\ref{main_sys}).} 
%one of the derivatives $\frac{\partial J}{\partial u^j_{n+m}}$ where $j=1,2,\ldots,N$ does not vanish identically and the following condition $D_x J = 0$ holds by virtue of system (\ref{main_sys}). }

Since the shift transformation $J\rightarrow D_nJ$ maps an $x$-integral into $x$-integral we can set here $k=0$.
Let us given a set of $x$-integrals $J^i (x, n, u_{n}, \ldots, u_{n + m_i})$ of the orders $m_i$. We call the set $J^1, \ldots, J^N$ a set of independent $x$-integrals if a system of functions $D^j_n J^i$, where $i=1,\ldots,N$, $j = 0,1,\ldots, s-s_i$ and $s = max\{s_i\}$ is functionally independent.

%We first slightly change the integrals by shifting the discrete argument
%\begin{equation*}
%W^i (x,n,u_n, \ldots, u_{n + s_i}) = D^{k_i}_n J^i(x, n, u_{n-k_i}, \ldots, u_{n+m_i}).
%\end{equation*}
%We denote $s = max_i s_i$. 

For the convenience of further work, we introduce the concept of a set of $n$-integrals of minimal orders: $n_1 < n_2 < \ldots < n_s$. Below we will give the definition of the notion of set of integrals of minimal orders. We will discuss a practical way of constructing this set by using the characteristic algebra in the proof of the Theorem~3.3, and also illustrate it with concrete examples (see \S~4).
Let
\begin{equation} \label{x1}
\omega = \omega(x, n, u_n, u_{n,x}, \ldots, u_{n,[k]})
\end{equation}
be a function satisfying the conditions
\begin{equation}	\label{x12}
D_n \omega = \omega, \quad \frac{\partial \omega}{\partial u_{n, [k]}} \neq 0.
\end{equation}
Minimality of the order $n_1$ means that a function $\omega$ of the form (\ref{x1}) satisfying (\ref{x12}), is a trivial integral as soon as $k < n_1$. Moreover, for $k = n_1$, $\omega$ is not a trivial integral. We denote $\omega^{1,1}, \ldots, \omega^{1, m_1}$ the complete set of functionally independent integrals of order $n_1$. Then we look for a number $n_2$ such that there is no integral of the order $n<n_2$, which cannot be represented as a function of $x$, the already found integrals $\omega^{1,1}, \ldots, \omega^{1, m_1}$ and their derivatives. Such an essentially new integral exists for $n = n_2$. In other words we look for a number $n_2$, assuming that $\omega$ is a function only on $x$, functions $\omega^{1,1}, \ldots, \omega^{1, m_1}$ and on the derivatives of these functions with respect to $x$, as soon as $k \in [n_1, n_2)$. We denote by $\omega^{2,1}, \ldots, \omega^{2, m_2}$ a functionally independent set of integrals of order $n_2$, which are not expressed in terms of $x, \omega^{1,1}, \ldots, \omega^{1, m_1}$ and their derivatives. Continuing this process, we obtain a set of $n$-integrals of minimal orders $n_1 < n_2 < \ldots < n_s$:
\begin{equation} \label{x3}
\omega^{1,1}, \ldots, \omega^{1, m_1}, \omega^{2,1}, \ldots, \omega^{2, m_2}, \ldots, \omega^{s,1}, \ldots, \omega^{s, m_s},
\end{equation}
where $m_1 + m_2 + \cdots + m_s = N$. 

In a similar way we determine a set of $x$-integrals of minimal orders.

Let system (\ref{main_sys}) have a set of $n$-integrals $I^1, I^2, \ldots I^N$ of the minimal orders  $s_1, s_2, \ldots s_N$ correspondingly. Let us denote $s= \mathrm{max}_j (s_j)$, $j=1,2,\ldots,N$ and determine a new set of integrals by taking $W^i = D^{s-s_i}_x I^i$. It is clear that functions $W^1, W^2, \ldots W^N$ are integrals of system (\ref{main_sys}) of one and the same order $s$.

{\bf Definition 2.3.}
{\it We say that $n$-integrals $I^1, I^2, \ldots I^N$ of system (\ref{main_sys}) constitute a set of independent in principal integrals if the following condition holds}
\begin{equation} \label{int_det}
\mathrm{det}\,\left( \frac{\partial W^k}{\partial u^i_{[s]}} \right)  \neq 0, \quad k =1,2,\ldots N.
\end{equation}

%In a similar way we call the set of $x$-integrals $J^1, J^2, \ldots J^N$ essentially independent if ???????????

Let system (\ref{main_sys}) have a set of $x$-integrals $J^1, J^2, \ldots J^N$ of the form\\ $J^i = J^i (x, n, u_{n}, \ldots, u_{n + m_i})$ of the minimal orders $r_1\leq r_2, \ldots, \leq r_N$ correspondingly. We determine a new set of integrals by taking $Z^i = D^{r_N-r_i}_n J^i$. 

{\bf Definition 2.4.}
{\it We say that $x$-integrals of the minimal orders $J^1, J^2, \ldots J^N$ of system (\ref{main_sys}) constitute a set of independent in principal integrals if the following condition holds}
\begin{equation} \label{int_det2}
\mathrm{det}\,\left( \frac{\partial Z^k}{\partial u^i_{n+r_N}} \right)  \neq 0, \quad k=1,2,\ldots, N.
\end{equation}

The notion of the independence in principal allows one to deduce an effective criterion of the independent set of integrals.

Let us consider a set of $n$-integrals
\begin{equation}  \label{x5}
\omega^1, \omega^2, \ldots, \omega^N
\end{equation}
of the minimal orders $k_1 \leq k_2 \leq \ldots \leq k_N$. We assume that (\ref{x5}) are analytic functions of the dynamical variables $u_n, u_{n,x}, u_{n,xx}, \ldots$ defined on a domain $D$.

{\bf Lemma 2.1.} {\it Set of $n$-integrals (\ref{x5}) is independent if and only if it is independent in principal.}

{\bf Proof.} Assume that integrals (\ref{x5}) are independent and the determinant vanishes
\begin{equation}  \label{x6}
\mathrm{det}\,\left( \frac{\partial W^k}{\partial u^i_{n,[k_N]}} \right)=
\left|  
\begin{array}{cccc}
\frac{\partial W^1}{\partial u^1_{[n,k_N]}} & \frac{\partial W^1}{\partial u^2_{[n,k_N]}} & ... & \frac{\partial W^1}{\partial u^N_{[n,k_N]}} \\
... & ... & ... & ...\\
\frac{\partial W^N}{\partial u^1_{[n,k_N]}} & \frac{\partial W^N}{\partial u^2_{[n,k_N]}} & ... & \frac{\partial W^N}{\partial u^N_{[n,k_N]}} 
\end{array}
 \right| = 0.
\end{equation}
The due to the theorem on implicit functions one of the functions $W^1, W^2, \ldots, W^N$ is expressed in terms of the others and the variables $x,n,u_n, u_{n,x}, \ldots, u_{n, [k_N - 1]}$. For the definiteness we set
\begin{equation}	\label{x7}
W^N = F(W^1, W^2, \ldots, W^{N-1}, x, n, u_{n}, u_{n,x}, \ldots, u_{n,[k_N-1]}).
\end{equation}
Let us expand the function (\ref{x7}) in the Taylor series around some fixed point $(W^1_0, W^2_0, \ldots, W^{N-1}_0) \in D$:
\begin{eqnarray} 
\fl F = \sum^{\infty}_{i_1, i_2, \ldots, i_{N-1} = 0} \alpha_{i_1 i_2 \ldots i_{N-1}} (x,n, u_n, u_{n,x}, \ldots, u_{n, [k_N-1]}) \times \nonumber\\
\times (W^1 - W^1_0)^{i_1}\cdots (W^{N-1} - W^{N-1}_0)^{i_{N-1}}. \label{x8}
\end{eqnarray}
Apply the operator $D_n - 1$ to both sides of (\ref{x8}) and get
\begin{equation*}
(D_n - 1) \alpha_{i_1 \ldots i_{N-1}} (x,n, u_n, u_{n,x}, \ldots, u_{n, [k_N - 1]}) = 0.
\end{equation*}
Therefore the coefficients of the series (\ref{x8}) are integrals of the orders less than $k_N$. Due to the definition of the set of integrals of the minimal orders they are expressed in terms of $x,\omega^1, \omega^2, \ldots, \omega^N$. Hence by virtue of the coincidence $W^N = \omega^N$ we obtain that integrals (\ref{x5}) are dependent, that contradicts the assumption above.

We prove now the converse statement. Assume that condition 
\begin{equation*}  
\left|  
\begin{array}{cccc}
\frac{\partial W^1}{\partial u^1_{n,[k_N]}} & \frac{\partial W^1}{\partial u^2_{n,[k_N]}} & ... & \frac{\partial W^1}{\partial u^N_{n,[k_N]}} \\
... & ... & ... & ...\\
\frac{\partial W^N}{\partial u^1_{n,[k_N]}} & \frac{\partial W^N}{\partial u^2_{n,[k_N]}} & ... & \frac{\partial W^N}{\partial u^N_{n,[k_N]}} 
\end{array}
 \right| \neq 0
\end{equation*}
holds and integrals $\omega^1, D_x \omega^1, \ldots, D^{k_n - k_1}_x \omega^1, \omega^2, D_x \omega^2, \ldots, D^{k_N - k_2}_x \omega^2, \ldots, \omega^N$ are dependent. The latter means that there exists a function $\Phi$ such that
\begin{equation}  \label{x9}
\Phi(\omega^1, \ldots, \omega^N, D_x \omega^1, \ldots, D_x \omega^N, \ldots, W^1, \ldots, W^N) = 0.
\end{equation}
We assume that function $\Phi$ satisfies the relation
\begin{equation} \label{x10}
\sum^N_{k=1} \left| \frac{\partial \Phi}{\partial W^k} \right|^2 \neq 0,
\end{equation}  
i.e. at least one of the derivatives $\frac{\partial \Phi}{\partial W^k}$ doesn't vanish identically.  If it is not the case and $\Phi$ doesn't depend on any of $W^j$ we apply to (\ref{x9}) operator $D_x$ and get a new relation of the form (\ref{x9}), that would satisfy (\ref{x10}). If necessary we perform this trik several times to get equation (\ref{x9}) satisfying (\ref{x10}). Then we differentiate the relation (\ref{x9}) with respect to each of the variables $u^i_s$, where $s=n_N$ and $i = 1,2,\ldots, N$. As a result we get an equation
\begin{equation*}
\frac{\partial \Phi}{\partial W^1} \frac{\partial W^1}{\partial u^i_s} + \frac{\partial \Phi}{\partial W^2} \frac{\partial W^2}{\partial u^i_s}  + \cdots + \frac{\partial \Phi}{\partial W^N} \frac{\partial W^N}{\partial u^i_s} = 0
\end{equation*}
which implies that
\begin{equation*}
\mathrm{det} \left( \frac{\partial W^k}{\partial u^i_s} \right) = 0.
\end{equation*}
Lemma is proved.

 For the $x$-integrals we can prove the analogue of the Lemma 1.1, claiming that set of $x$-integrals is independent if and only if it is independent in principal.

Now we will prove that a set of $n$-integrals independent in principal constitutes in a sense a basis of integrals by showing that any integral of the system is expressed through the members of this set.

{\bf Theorem 2.1.}
{\it Assume that system (\ref{main_sys}) has an independent in principal set of $n$-integrals $I^1, \ldots, I^N$ analytically depending on the dynamical variables in a domain $D$, having the minimal orders $n_1 \leq n_2 \leq ... \leq n_N$. Then any other $n$-integral $G$ of (\ref{main_sys}) analytic in $D$ is a function of}
\begin{equation*}
x, I^1, I^2, \ldots, I^N, D_x I^1, D_x I^2, \ldots D_x I^N, D^2_x I^1, D^2_x I^2, \ldots, D^2_x I^N, \ldots 
\end{equation*}

{\bf Proof.} Denote by $s$ the order of the integral $G$. If $s \leq n_N$ then the statement of the theorem follows from the definition of the set of the minimal orders. For the simplicity we take the case $s = n_N + 1$. Then by applying operator $D_x$ we get a new set of the integrals 
$W^1_1, W^2_1, \ldots, W^N_1$ where $W^i_1 := D_x W^i$. Since the relations hold
\begin{equation} \label{y1}
\mathrm{det} \left| \frac{\partial W^i_1}{\partial u^i_{n, [s]}} \right| = \mathrm{det} \left| \frac{\partial W^i}{\partial u^i_{n, [s-1]}} \right| \neq 0
\end{equation}
function $G$ can be represented as
\begin{equation}  \label{y2}
G = F(W^1_1, \ldots, W^N_1, u_{n,[s]}, u_{n, [s-1]}, \ldots, u_n, x, n).
\end{equation}
Moreover we can get rid $u_{n,[s]}$ due to (\ref{y1}) and rewrite (\ref{y2}) as
\begin{equation}	\label{y3}
G = \bar{F}(W^1_1, \ldots, W^N_1, W^1, \ldots, W^N, u_{n,[s-1]}, \ldots, u_n, x, n).
\end{equation}
Then we expand $\bar{F}$ into a power series around a point $(W^1_0, \ldots, W^N_0; W^1_{01}, \ldots, W^N_{01}) \in D:$
\begin{eqnarray*}
\bar{F} = \sum \alpha_{i_1 \ldots i_N, j_1, \ldots, j_N} (x,n,u_n, u_{n,x}, \ldots, u_{n, [s-1]}) \times \\
\times (W^1 - W^1_0)^{i_1} \cdots (W^N - W^N_0)^{i_N} (W^1_1 - W^1_{01})^{j_1} \cdots (W^N_1 - W^N_{01})^{j_N}.
\end{eqnarray*}
Now we apply operator $D_n - 1$ to $\bar{F}$ and prove that coefficients of the series are $n$-integrals of the order less than or equal to $s-1 = n_N$, therefore they are expressed through $x$, integrals $I^1, I^2, \ldots, I^N$ and their derivatives with respect to $x$. This completes the proof.

A version of the Theorem 2.1 concerned the $x$-integrals is also true. It can be  proved in a similar way.

{\bf Theorem 2.2.}
{\it Assume that system (\ref{main_sys}) has an independent in principal set of $x$-integrals $J^1, \ldots, J^N$ analytically depending on the dynamical variables in a domain $\bar{D}$, having the minimal orders $n_1 \leq n_2 \leq ... \leq n_N$. Here for any $j$ integral $J^j$ is of the form $J^j = J^j (x,n, u_n, \ldots, u_{n+n_j})$.  Then any other $x$-integral $\bar{G} = \bar{G}(x, n, u_n, \ldots, u_{n+s})$ of (\ref{main_sys}) analytic in $\bar{D}$ is a function of}
\begin{equation*}
n, J^1, J^2, \ldots, J^N, D_n J^1, D_n J^2, \ldots D_n J^N, D^2_n J^1, D^2_n J^2, \ldots, D^2_n J^N, \ldots 
\end{equation*}

{\bf Definition 2.5.} 
{\it System (\ref{main_sys}) is called Darboux integrable if it has sets of integrals  independent in principal on both characteristic directions of $x$ and $n$.}

\section{Algebraic criterion of Darboux integrability}

In this section we introduce characteristic Lie-Rinehart algebras for the system of differential-difference equations of the form (\ref{main_sys}). These objects are very important attributes of the Darboux integrability since the finite dimensionality property of the characteristic algebras is necessary and sufficient for the Darboux integrability. For the scalar discrete equation the notion of the characteristic algebra was introduced for the first time in \cite{sigma2005}. For the scalar differential-difference equation it was studied in \cite{HPZ2009}. The notion was generalized to the systems of differential-difference equations in \cite{HabibullinKhakimova}. Below we use the definitions given in  \cite{HabibullinKhakimova}.

%In this section we give necessary and sufficient conditions of Darboux integrability in terms of the characteristic Lie-Rinehart algebras. Characteristic algebra provides an efficient tool for deriving conditions of Darboux integrability for system \eqref{main_sys} and thus due to the reduction method for equations with three independent variables \eqref{eq0} as well. Let us emphasize that from the classification viewpoint the necessary integrability conditions are very important.

Suppose that system (\ref{main_sys}) has an integral in the direction of $ x $, we denote it by $J$. It is easy to check that function $J$ can  depend only on the variables $ x, n, u_n, u_ {n \pm 1}, u_ {n \pm 2}, \ldots $. According to Definition 2.2, we have
%In order to find the integrals one can use the so-called characteristic operators. Let us discuss how these operators are derived. Assume that $J$ is an $x$-integral, then according to the definition we have $D_x J(x,n,u_n,u_{n\pm1}, u_{n\pm2},\ldots)=0$, we use the chain rule and get
\begin{equation}\label{KJ}
\fl D_x J = K_0J=\left(\frac{\partial}{\partial x}+\sum_{j=1}^{N} u_{n,x}^j \frac{\partial}{\partial u_{n}^j}+u_{n+1,x}^j\frac{\partial}{\partial u_{n+1}^j}+ u_{n-1,x}^j \frac{\partial}{\partial u^j_{n-1}} +\ldots\right)J=0.
\end{equation}
%Let's  specify the operator $K_0$ by means of the system \eqref{main_sys} and obtain
Taking into account system (\ref{main_sys}), the operator can be easily rewritten in the form
\begin{equation}\label{K0}
\fl K_0=\frac{\partial}{\partial x}+\sum_{j=1}^{N}(u_{n,x}^j \frac{\partial}{\partial u_{n}^j}+F_{n}^j\frac{\partial}{\partial u_{n+1}^j}+ G_{n}^j \frac{\partial}{\partial u^j_{n-1}} +F_{n+1}^j\frac{\partial}{\partial u_{n+2}^j}+ G_{n-1}^j \frac{\partial}{\partial u^j_{n-2}} +\ldots)
\end{equation}
where $F^j_{n+i}=F^j(x,n+i,u_{n+i,x}^j,u_{n+i},u_{n+i+1})$ and\\ $G^j_{n-i}= G^j(x,n-i,u_{n-i,x}^j,u_{n-i},u_{n-i-1})$. We call $K_0$ the characteristic operator in the $x$ direction.

Consequently, the $x$ -integral is a solution to the first order linear differential equation
\begin{equation}  \label{eq3_02}
K_0 J = 0.
\end{equation}

The peculiarity of the equation is that the coefficients of the equation depend on the derivatives of the function $u$ with respect to $x$, while the solution $J$ cannot depend on these variables. This actually means that equation (\ref{eq3_02}) is strongly overdetermined. Indeed, the sought  function, in addition to (\ref{eq3_02}), satisfies several more equations
%An important feature of the equation (\ref{KJ}) is that the solution $J$ does not depend on the variables $u^1_{n,x}, u^2_{n,x},\ldots,u^N_{n,x}$ despite the fact that the coefficients of the equation depend on them. Therefore in addition to (\ref{KJ}) any $x$-integral solves also a set of the following equations
\begin{equation}\label{Xk}
X_kJ=0, \quad k=1,2,\ldots,N,
\end{equation}
where $X_k=\frac{\partial}{\partial u^k_{n,x}}$. Thus any $x$-integral is nullified by the operators
\begin{equation}\label{XkK}
X_1, X_2, \ldots, X_N, K_0
\end{equation}
their multiple commutators as well as by linear combinations of the obtained operators with variable coefficients, depending on the dynamical variables. Therefore the $x$-integral belongs to the kernel of the operators which lie in some generalization of the Lie algebra, where the operators can be multiplied  not only by constants but also by elements of the ring of locally analytic functions depending on the dynamical variables, we denote the ring by $A$.

In other words, any $x$-integral belongs to the kernel of the operators in the Lie-Rinehart algebra $L_x$, generated by the operators (\ref{XkK}) over the ring $A$. We call it characteristic algebra in the direction of $x$. Note that the operation of multiplication by a function and taking the commutator of two operators are connected by conditions:
\begin{itemize}
\item[1)] $[W_1,aW_2]=W_1(a)W_2+a[W_1,W_2]$,
\item[2)] $(aW_1)b=aW_1(b)$,
\end{itemize} 
that are valid for any $W_1,W_2\in L_x$ and $a,b\in A$. This means that, if $W_1\in L_x$ and $a\in A$ then $aW_1\in L_x$ (see \cite{Rinehart}). 

The algebra $L_x$ is finite dimensional if there exists a  basis having a finite number of the elements $W_1,W_2,\ldots,W_k\in L_x$ such that an arbitrary element $W\in L_x$ can be represented as a linear combination of the form
\begin{equation}	\label{linear_comb}
W=a_1W_1+a_2W_2+\dots +a_kW_k,
\end{equation}
where the coefficients are functions $a_1, a_2, \ldots, a_k\in A$. If in (\ref{linear_comb}) $W=0$ then we have $a_1=0$, $a_2=0$, $\ldots$, $a_k=0$.

The requirement of local analyticity of the elements of the ring A is important from the point of view of the existence of a basis.

Let us briefly discuss the difference between Lie and Lie-Rinehart algebras.

{\bf Example 3.1.} Obviously, the Lie algebra generated by the operators $Z_1 = x^2 \frac{\partial}{\partial x}$ and $Z_2 = x^3 \frac{\partial}{\partial x}$ is infinite-dimensional. For example, the commutator $[Z_1,Z_2]=x^4\frac{\partial}{\partial x}$ is not a linear combination of $Z_1$ and $Z_2$ with constant coefficients.  At the same time the Lie-Rinehart algebra corresponding to the ring $A$ of functions analytic in the domain $x \neq 0$ of the complex plane generated by the same operators is one-dimensional, since any element $Z$ in the algebra can be represented as $Z = f(x) Z_0$, where $Z_0 = \frac{\partial}{\partial x}$, since here linear combinations with variable coefficients are allowed.

The algebraic background of the Darboux integrability theory is based on the following important fact.

{\bf Theorem 3.1.}
{\it System (\ref{main_sys}) admits an independent set of $x$-integrals if and only if its characteristic Lie-Rinehart algebra $L_x$ has a finite dimension.}

The proof of Theorem 3.1. for the case $N=2$ is given in the recent article \cite{ZhKuz21} by A.V.Zhiber and one of the authors. The proof, presented in \cite{ZhKuz21}, carries over without special difficulty to the general case, therefore we do not present it here.

Let us turn to the consideration of $ n $-integrals. Suppose that $I $ is a nontrivial $n $-integral of system (\ref{main_sys}). It is easy to check that $I$ can depend only on $n $, $x $ and dynamical variables $u_n, u_{n,x}, u_{n,xx},\ldots$. According to Definition 2.1, the equality holds
\begin{equation*}
I(x,n+1,u_{n+1},u_{n+1,x},u_{n+1,xx},...)=I(x,n,u_n,u_{n,x},u_{n,xx},\ldots),
\end{equation*}
which, by virtue of equation (\ref{main_sys}), can be rewritten in the following form
\begin{equation} \label{DnI}
I(x,n+1,u_{n+1},F,D_xF, D_x^2F,\ldots)=I(x,n,u_n,u_{n,x},u_{n,xx},\ldots).
\end{equation}
Hence, it is clear that to find the $n$-integral it is necessary to solve a functional equation, since the equality (\ref{DnI}) connects the values of the required function at two different points. Note that (\ref{DnI})  admits a large number of differential consequences. Indeed, since the right-hand side of (\ref{DnI})  does not contain dependence on the variable $u_{n+1}$, then for any $j$ the equality $\frac{\partial}{\partial u^j_{n+1}} D_n I = 0$ holds, from which it easily follows that
\begin{equation} \label{derivative}
Y_{j,1}I:=D_n^{-1}\frac{\partial}{\partial u^j_{n+1}}D_nI=0\quad \mbox{for}\quad j=1,2,\ldots,N.
\end{equation}
In expanded form, (\ref{DnI}) is represented as a vector field
\begin{equation} \label{vectorfields}
Y_{j,1}=\frac{\partial}{\partial u^j_{n}}+\sum_{i=1}^N D_n^{-1}\left(\frac{\partial F_n^i}{\partial u^j_{n+1}}\right)\frac{\partial}{\partial u^i_{n,x}}
+D_n^{-1}\left(\frac{\partial F_{n,x}^i}{\partial u^j_{n+1}}\right)\frac{\partial}{\partial u^i_{n,xx}}+\dots,
\end{equation}
where $F_{n,x}^i:=D_xF_{n}^i$. We set $Y_{j,0}=\frac{\partial}{\partial u^j_{n+1}}$ and then rewrite representation as follows
\begin{equation*} \label{vectorfields1}
Y_{j,1}=\frac{\partial}{\partial u^j_{n}}+\sum_{i=1}^N D_n^{-1}\left(Y_{j,0}\left(F_n^i\right)\right)\frac{\partial}{\partial u^i_{n,x}}
+D_n^{-1}\left(Y_{j,0}\left(F_{n,x}^i\right)\right)\frac{\partial}{\partial u^i_{n,xx}}+\dots. 
\end{equation*}
Repeatedly applying the operator $D_n$ to the equality $D_n I = I$, we obtain $D^k_n I = I$, applying the operator $\frac{\partial}{\partial u^j_{n+1}}$ to the latter, we derive the equality
\begin{equation} \label{derivative-k}
Y_{j,k}I:=D_n^{-k}\frac{\partial}{\partial u^j_{n+1}}D_n^kI=0\quad \mbox{for}\quad j=1,2,\ldots,N.
\end{equation}
Operators $Y_{j,k}$ allow coordinate representation
\begin{equation} \label{vectorfieldsk}
Y_{j,k}=\sum_{i=1}^N D_n^{-1}\left(Y_{j,k-1}\left(F_n^i\right)\right)\frac{\partial}{\partial u^i_{n,x}}
+D_n^{-1}\left(Y_{j,k-1}\left(F_{n,x}^i\right)\right)\frac{\partial}{\partial u^i_{n,xx}}+\dots. 
\end{equation}
Since the coefficients of the equations (\ref{derivative-k}) depend on the variables $u^j_{n-s}$ for $1\leq s\leq k$ and $1\leq j\leq N$, while the solution to the equation does not depend on them, then such equations 
\begin{equation*} \label{barX}
\bar X_{i,s}I=0,
\end{equation*}
where $\bar X_{i,s}=\frac{\partial}{\partial u^i_{n-s}}$ should hold for these values of $i$ and $s$.

In what follows $A_k$ denotes a ring of locally analytic functions of the variables $u_{n-k},u_{n-k+1},\ldots,u_n$; $u_{nx},u_{n,xx},u_{n,xxx},\ldots$. In the next theorem, we define the characteristic Lie-Rinehart algebra  $L_n$ and derive a necessary condition for the existence of $n$-integrals.

{\bf Theorem 3.2.}
{\it If system (\ref{main_sys}) admits an independent set of $n$-integrals then the following two conditions hold:

1) The linear space  $V$ spanned by the operators $\{ Y_{i,s}\}$ has finite dimension, which we denote by $N_1$. Assume that $Z_1, Z_2,\ldots,Z_{N_1}$ constitute a basis in $V$, such that for any $Z\in V$ we have an expansion
\begin{equation*}
Z=\lambda_1Z_1+\lambda_2Z_2+\ldots+\lambda_{N_1}Z_{N_1},
\end{equation*}
We emphasize that the coefficients in this expansion, as a rule, are not constant, they are analytic functions of dynamic variables. Then due to construction of the space $V$ there exists a number $N_2$, such that $[Z_j,\bar X_{i,s}]=0$ for all $j=1,2,\ldots,N_1$, $i=1,2,\ldots,N$ and $s>N_2$.

2) The Lie-Rinehart algebra $L_n$ generated by the operators $\{Z_j\}_{j=1}^{N_1}$ and $\{\bar X_{i,s}\}_{s=1,i=1}^{N_2,N}$ over the ring $A_{N_2}$ has a finite dimension.}

%The proof of these two theorems is beyond the scope of this paper. In the particular case when $N=1$ Theorems \ref{Th1}, \ref{Th2} are proved in \cite{HZS2010}. For systems of differential equations of hyperbolic type, similar statement is proved in \cite{Zhiber2007}. 

Proof of the theorem. Assume that system (\ref{main_sys}) admits an essentially independent set of $n$-integrals $I^{(1)}$, $I^{(2)}$,... , $I^{(N)}$ of the order $m$. Then the relation holds
\begin{equation} \label{Jacobian}
\det\left( \frac{\partial I}{\partial u_{[m]}} \right)\neq0
\end{equation}
where $I=(I^{(1)}, I^{(2)},..., I^{(N)})$. Our goal is now to prove that the corresponding characteristic algebra is of finite dimension.

Let us change the variables in the characteristic vector fields $Y_{j,k}$ in the following way.
We pass from the subset of the dynamical variables 
$$u_n, u_{n,x}, u_{n,xx},... $$
to the subset 
$$u_n, u_{n,x}, u_{n,xx},...,u_{n,[m-1]},   I,D_x(I), D^2_x(I),\ldots. $$
In terms of these new variables characteristic operators take the form
$$Y_{j,k}=\sum_i a^{(0)}_{j,k,i}\frac{\partial}{\partial u^i_{n}}+a^{(1)}_{j,k,i}\frac{\partial}{\partial u^i_{n,x}}+...+a^{(m)}_{j,k,i}\frac{\partial}{\partial I^i} +a^{(m+1)}_{j,k,i}\frac{\partial}{\partial I^i_{[1]}}+ ...$$
where $I^i_{[s]}=D_x^s(I^i)$. However, since the variables $I^i$, $I^i_{[1]}$,... are n-integrals we have $a^{(m+s)}_{j,k,i}=Y_{j,k}(I^i_{[s]})=0$ for all $s\geq0$. Therefore the characteristic algebra $L_n$ is generated by the differential operators 
$$Y_{j,k}=\sum_i a^{(0)}_{j,k,i}\frac{\partial}{\partial u^i_{n}}+a^{(1)}_{j,k,i}\frac{\partial}{\partial u^i_{n,x}}+...+a^{(m-1)}_{j,k,i}\frac{\partial}{\partial u^i_{n,[m-1]}} $$
and a finite set of the operators $X_{j,k}$. Obviously this algebra is of a finite dimension. Consequently both conditions 1) and 2) of the Theorem 3.2. are satisfied.

%%%%%%%%%%%%%%%%%%%%%%%%%%%%%%%%%%%%%%%%%%%%%%%%%%%%%%%%%%%%%%%%%%%%%

The following theorem gives a sufficient condition for the existence of the $n$-integrals. Theorems 3.2, 3.3 were given without proofs in work \cite{HabibullinKhakimova}. In the course of the proof, we clarified the essence the concept of a complete set of integrals. 

{\bf Theorem 3.3.}
{\it If algebra $L_n$ for the system (\ref{main_sys}) is of a finite dimension then the system admits a set of independent $n$-integrals.}

{\bf Proof.}
The proof consists of two essentially different parts. In the first part we construct preliminary set of functions which are  independent in principal solutions to the characteristic system of equations (see below (\ref{p1})). In part~2  we construct the desired set of integrals by using the preliminary set found in the first part. This complexity is due to the specificity of discrete equations.

{\bf Part 1.} Let us assume that algebra $L_n$ is of a finite dimension. We denote by $L$ a linear span of $\bar{X_{i,s}}$. We first decompose algebra $L_n$ into a direct sum of two subalgebras $L_n = L \oplus B$. Then we focus on the subalgebra $B$, consisting of elements of the following form:
\begin{eqnarray}
&Z_i = \frac{\partial}{\partial u^i_n} + \sum^N_{i=1} \left( \alpha^i_{(1)} \frac{\partial}{\partial u^i_{n,x}} + \beta^i_{(1)} \frac{\partial}{\partial u^i_{n,xx}}+\cdots \right), \quad i=1,2,\ldots, N,\\ \label{linsys}
&Z_{N+k} = \sum^N_{i=1} \left( \gamma^i_{(N+k)} \frac{\partial}{\partial u^i_{n,x}} + \delta^i_{(N+k)} \frac{\partial}{\partial u^i_{n,xx}} \right), \quad k \geq 1.
\end{eqnarray}
The aim of this part is to construct an independent in principal set of solutions of the system 
\begin{equation}\label{p1}
Z_j(n)G(n)=0
\end{equation}
where the operators $Z_j$ are from the list (\ref{linsys}).

In the proof, we consider various  possible situations corresponding to the structure of the subalgebra $B$ and specify how to choose the solutions of the system (\ref{p1}) in order to obtain an independent in principal set of solutions. It is clear that the dimension of the subalgebra $B$ satisfies the inequality ${\rm dim}\, B \geq N+1$. 
Let us first assume that ${\rm dim}\, B = N+1$. Then the basis can be taken as the union of the vector fields $Z_i$, $i=1,2,\ldots N$ and 
\begin{equation}  \label{ZN1}
Z_{N+1} = \sum^N_{i=1} \left(  \varphi^i_{N+1,m} \frac{\partial}{\partial u^i_{n,[m]}} + \varphi^i_{N+1,m+1} \frac{\partial}{\partial u^i_{n,[m+1]}} + \cdots \right), \quad m \geq 1.
\end{equation}
If $m=1$ then we consider the system
\begin{equation} \label{eq3_3}
\fl \left( \frac{\partial}{\partial u^i_n} + \sum^N_{i=1} \alpha^i_{(1)} \frac{\partial}{\partial u^i_{n,x}} \right) \omega = 0, \quad \sum^N_{k=1} \varphi^k_{N+1,1} \frac{\partial \omega}{\partial u^k_{n,x}} = 0, \quad i=1,2,\ldots N.
\end{equation}
This system consists of $N+1$ equations for a function of $2N$ variables $u^i_n, u^i_{n,x}$, $i=1,\ldots,N$. Thus it
has $N-1$ functionally independent solutions $\omega^i (x,n, u_n, u_{n,x})$, $i=1,2,\ldots,N-1$. 

Now let us consider the system:
\begin{eqnarray}
& \left( \frac{\partial}{\partial u^i_n} + \sum^N_{i=1} \left( \alpha^i_{(1)} \frac{\partial}{\partial u^i_{n,x}} + \beta^i_{(1)} \frac{\partial}{\partial u^i_{n,xx}} \right) \right) W = 0, \quad i=1,2,\ldots,N,\nonumber\\
& \sum^N_{k=1} \left( \varphi^k_{N+1,1} \frac{\partial}{\partial u^k_{n,x}} + \varphi^k_{N+1,2} \frac{\partial}{\partial u^k_{n,xx}}\right) W = 0. \label{eq3_4}
\end{eqnarray}
It consists of $N+1$ equations on a function of $3N$ variables. Thus we have $2N-1$ functionally independent solutions
\begin{equation*}
\fl \omega^i(x,n,u_{n}, u_{n,x}), \quad D_x \omega^i(x,n,u_{n}, u_{n,x}), \quad W(x,n,u_n, u_{n,x}, u_{n,xx}), \quad i=1,2,\ldots, N-1.
\end{equation*}
Thus system (\ref{p1}) possesses a complete set of independent in principal solutions $\omega^i$, $W$, $i=1,2,\ldots, N-1$.

Let us consider case $m \geq 2$. Then to find integrals of the first order $\omega(x,n,u_n, u_{n,x})$ we have the following system:
\begin{equation*}
\left( \frac{\partial}{\partial u^i_n} + \sum^N_{i=1} \alpha^i_{(1)} \frac{\partial}{\partial u^i_{n,x}} \right) \omega = 0, \quad i=1,2,\ldots,N
\end{equation*}
with $N$ equations and $2N$ independent variables. Thus the system admits $N$ functionally independent solutions $\omega^i(x,n,u_n, u_{n,x})$, $i=1,2,\ldots,N$.

Now we consider the case when ${\rm dim} \, B > N+1$. Assume that ${\rm dim}\, B = N+2$, then the basis is defined by the vector fields $Z_i$, $i=1,2,\ldots, N$ and 
\begin{eqnarray*}
&\fl Z_{N+1} = \sum^N_{i=1} \left( \varphi^1_{N+1,m} \frac{\partial}{\partial u^i_{n,[m]}} + \varphi^i_{N+1,m+1} \frac{\partial}{\partial u^i_{n,[m+1]}} + \cdots \right),\\
&\fl Z_{N+2} = \sum^N_{i=1} \left( \psi^i_{N+2,s} \frac{\partial}{\partial u^i_{n,[s]}} + \psi^i_{N+2,s+1} \frac{\partial}{\partial u^i_{n,[s+1]}} + \cdots \right), \quad m,s \geq 1, \quad s \geq m.
\end{eqnarray*}

If $m \geq 2$ then the system of equations
\begin{equation*}
\left( \frac{\partial}{\partial u^i_n} + \sum^N_{i=1} \alpha^i_{(1)} \frac{\partial} {\partial u^i_{n,x}} \right) \omega = 0, \quad i=1,2,\ldots,N
\end{equation*}
has $N$ functionally independent solutions $\omega^i = \omega^i(x,n, u_{n}, u_{n,x})$ because here we have $N$ equations and $2N$ variables. These solutions also satisfy equations $Z_{N+1} \omega^i = 0$, $Z_{N+2} \omega^i = 0$.

Let us consider the case when $m=1$. If $s \geq 3$ then in addition to system (\ref{eq3_4}) we have also the following equation
\begin{equation*}
\fl \tilde{Z}_{N+2} = \psi^1_{N+2,3} \frac{\partial}{\partial u^1_{n,xxx}} + \ldots + \psi^N_{N+2,3} \frac{\partial}{\partial u^N_{n,xxx}} + \sum^{\infty}_{k=4} \left( \mu^1_k \frac{\partial}{\partial u^1_{n,[k]}} + \ldots + \mu^N_k \frac{\partial}{\partial u^N_{n,[k]}} \right).
\end{equation*}
As it was shown earlier, the entire family of solutions of system (\ref{eq3_4}) is given by the integrals
\begin{equation*}
\omega^i(x,n,u_n,u_{n,x}), \quad W(x,n,u_n, u_{n,x}, u_{n,xx}), \quad i = 1,2,\ldots, N-1.
\end{equation*}
In addition the integrals satisfy also equations $\tilde{Z}_{N+2} \omega^i=0$,  $\tilde{Z}_{N+2} W = 0$, $ i = 1,2,\ldots, N_1$. Thus these integrals define the whole family of solutions of system (\ref{eq3_4}).

Let us assume that $s = 2$ (if $m=1$). Then we obtain system (\ref{eq3_3}) for unknown $\omega(x,n,u_n,u_{n,x})$. So this system has $N-1$ functionally independent integrals.
To find solutions of the form $W(x,n,u_n, u_{n,x}, u_{n,xx})$ we obtain system (\ref{eq3_4}) and additionally the equation
\begin{equation*}
 \sum^N_{i=1} \psi^i_{N+2,2} \frac{\partial W}{\partial u^i_{n,xx}} = 0.
\end{equation*}
Thus we deal with the system of $N+2$ equations on a function on $3N$ variables. In this case the system has $2 N - 2$ functionally independent solutions $\omega^i(x,n,u_n, u_{n,x})$, $D_x \omega^i(x,n,u_n,u_{n,x})$. To find solutions of the form $I(x,n, u_n, u_{n,x}, u_{n,xx}, u_{n,xxx})$ we consider the system:
\begin{eqnarray*}
&\left( \frac{\partial}{\partial u^i_n} + \sum^N_{i=1} \left( \alpha^i_{(1)} \frac{\partial}{\partial u^i_{n,x}} + \beta^i_{(1)} \frac{\partial}{\partial u^i_{n,xx}} + \gamma^i_{(1)} \frac{\partial}{\partial u^i_{n,xxx}} \right) \right) I =0, \quad i=1,2,\ldots,N,\\
&\sum^N_{k=1} \left( \varphi^k_{N+1,1} \frac{\partial}{\partial u^i_{n,x}} + \varphi^k_{N+1,2} \frac{\partial}{\partial u^i_{n,xx}} + \varphi^k_{N+1,3} \frac{\partial}{\partial u^i_{n,xxx}} \right) I = 0,\\
&\sum^N_{i=1} \left( \psi^i_{N+2,2} \frac{\partial}{\partial u^i_{n,xx}} + \psi^i_{N+2,3} \frac{\partial}{\partial u^i_{n,xxx}} \right) I = 0.
\end{eqnarray*}
This is a system of $N+2$ equations on a function on $4N$ variables. Thus, it has $3N-2$ solutions: 
\begin{equation*}
\omega^i(x, n, u_n, u_{n,x}),\, D_x \omega^i(x, n, u_n, u_{n,x}), \, D^2_x \omega^i (x,n, u_n, u_{n,x}), \, I(x,n, u_n, u_{n,x}, u_{n,xx}, u_{n,xxx}),
\end{equation*}
$i=1,2, \ldots, N-1$.

It remains to consider the case $s=1$ (if $m=1$). We can assume that vector fields $Z_{N+1}$, $Z_{N+2}$ have the forms:
\begin{eqnarray}
&\fl \tilde{Z}_{N+1} = \frac{\partial}{\partial u^1_{n,x}} + \varphi^2_1 \frac{\partial}{\partial u^2_{n,x}} + \cdots + \varphi^N_1 \frac{\partial}{\partial u^N_{n,x}} + \sum^{\infty}_{l=2} \left( d^1_l \frac{\partial}{\partial u^1_{n,[l]}} + \cdots + d^N_l \frac{\partial}{\partial u^N_{n,[l]}} \right), \label{new1}\\
&\fl \tilde{Z}_{N+2} = s^1_1 \frac{\partial}{\partial u^1_{n,x}} + \cdots + s^N_1 \frac{\partial}{\partial u^N_{n,x}} + \sum^{\infty}_{l=2} \left( \mu^1_l \frac{\partial}{\partial u^1_{n,[l]}} + \cdots + \mu^N_l \frac{\partial}{\partial u^N_{n,[l]}} \right). \nonumber
\end{eqnarray}
If
\begin{equation}	\label{eq3_6}
\frac{s^1_1}{1} = \frac{s^2_1}{\varphi^2_1} = \ldots = \frac{s^N_1}{\varphi^N_1}
\end{equation}
then we replace $\tilde{Z}_{N+2}$ by
\begin{equation*}
\tilde{\tilde{Z}}_{N+2} = \tilde{Z}_{N+2} - s^1_1 \tilde{Z}_{N+1}.
\end{equation*}
This vector field has the following form:
\begin{equation*}
\tilde{\tilde{Z}}_{N+2} = \eta^1_k \frac{\partial}{\partial u^1_{n,[k]}} + \cdots + \eta^N_k \frac{\partial}{\partial u^1_{n,[k]}} + \sum^{\infty}_{s=k+1} \left( \nu^1_s \frac{\partial}{\partial u^1_{n,[s]}} + \cdots + \nu^N_s \frac{\partial}{\partial u^1_{n,[s]}} \right), \quad k \geq 2.
\end{equation*}
This case has already been investigated above.

If at least one of the equalities (\ref{eq3_6}) is not satisfied then the operator $\tilde{Z}_{N+2}$ can be replaced by
\begin{equation}	\label{eq3_7}
\tilde{\tilde{Z}}_{N+2} = \eta^2_1 \frac{\partial}{\partial u^2_{n,x}} + \cdots + \eta^N_1 \frac{\partial}{\partial u^N_{n,x}} + \sum^{\infty}_{s=2} \left( \nu^1_s \frac{\partial}{\partial u^1_{n,[s]}} + \cdots + \nu^N_s \frac{\partial}{\partial u^N_{n,[s]}} \right).
\end{equation}
So, we have the following system:
\begin{equation*}
Z_i W = 0, \quad \tilde{Z}_{N+1} W = 0, \quad \tilde{\tilde{Z}}_{N+2} W = 0, \quad i=1,2,\ldots, N,
\end{equation*}  
where $\tilde{Z}_{N+1}$ is defined by (\ref{new1}), $\tilde{\tilde{Z}}_{N+2}$ is defined by (\ref{eq3_7}). The system has $N-2$ solutions $\omega^i(x,n,u_n, u_{n,x})$, $i=1, \ldots, N-2$, because in this case we have $N+2$ equations and $2N$ variables. The system has $2N-2$ solutions: 
\begin{eqnarray*}
\omega^i(x,n, u_n, u_{n,xx}), \quad D_x \omega^i(x,n,u_n, u_{n,x}), \quad i=1,\ldots N-2, \\
W^1(x,n,u_n, u_{n,x}, u_{n,xx}), \quad W^2(x,n,u_n, u_{n,x}, u_{n,xx}).
\end{eqnarray*}
Thus we obtain a complete set of functions from the kernel of the operators in the algebra $L_n$ containing functions $\omega^i(x,n,u_n, u_{n,x}, u_{n,xx})$, $i=1,\ldots,N-2$, \\
$W^1(x,n,u_n,u_{n,x})$, $W^2(x,n,u_n,u_{n,x}, u_{n,xx})$. The case ${\rm dim} B > N+2$ is considered in a similar way. That completes the first part of the proof, where we constructed a set of functions defining an independent in principal set of solutions to the system (\ref{p1}).  However these solutions aren't $n$-integrals to the system of differential-difference equations (\ref{main_sys}) because they do not satisfy the shift condition (see Definition 2.1). 

{\bf Part 2.} Now we have to construct required integrals by using the obtained in the first part set of solutions to the first order linear equations of the form
\begin{equation} \label{pr3_eq2}
Z_j(n) G(n) = 0, \quad j=1,2,\ldots, M,\quad M=N+k,
\end{equation}
where the operators $Z_j=Z_j(n)$ are given in (\ref{linsys}). Denote these solutions as $G^{(1)}$, $G^{(2)}$,\ldots, $G^{(N)}$, recall that they constitute a set of independent in principal functions. Introduce a notation for the vector valued solution of the system (\ref{pr3_eq2}) by setting 
$$G_n = (G^{(1)}_n, G^{(2)}_n, \ldots, G^{(N)}_n).$$
We notice that for $\forall i$ function $G^i_n$ depends on $u_n, u_{n,x}, \ldots, u_{n,[m]}$ and satisfy the inequality
\begin{eqnarray}	
\frac{\partial G_n}{\partial u_{n,[m]}} = {\rm det} \left( 
\begin{array}{cccc}
\frac{\partial G^{(1)}}{\partial u^1_{n,[m]}} & \frac{\partial G^{(1)}}{\partial u^2_{n,[m]}} &\ldots &\frac{\partial G^{(1)}}{\partial u^N_{n,[m]}} \\
 \ldots & \ldots & \ldots & \ldots\\\
\frac{\partial G^{(N)}}{\partial u^1_{n,[m]}} & \frac{\partial G^{(N)}}{\partial u^2_{n,[m]}} & \ldots & \frac{\partial G^{(N)}}{\partial u^N_{n,[m]}}
\end{array}
 \right)\neq0.\label{pr3_eq3}
\end{eqnarray}

{\bf Lemma 3.1.} 
{\it Functions $\left\{ G_n \right\}^{n = +\infty}_{n = -\infty}$ solve one and the same system of the first order linear equations, i.e. the relations hold}
\begin{equation}	\label{pr3_eq4}
Z_j(0) G_n = 0, \quad j=1,2,\ldots, M.
\end{equation}

Proof of the Lemma. Due to the definition following relations hold
\begin{eqnarray}
& D^{-1} Y_{j,1} D = Y_{j,2} + X_{j,1},\nonumber\\
& D^{-1} Y_{j,k} D = Y_{j,k+1}, \quad k \geq 2,\label{pr3_eq5}\\
& D^{-1} X_{j,k} D = X_{j,k+1}. \nonumber
\end{eqnarray}
Relations (\ref{pr3_eq5}) imply that subalgebra $B$ of the Lie-Rinehart algebra $L_n$ generated by the operators $\left\{ Z_j(n) \right\}^N_{j=1}$ is invariant under the action of the automorphism defined due to the rule
\begin{equation}		\label{pr3_eq8}
Z \rightarrow D^{-1}_n Z D_n.
\end{equation}
Therefore if $G_n$ is a solution of the system (\ref{pr3_eq4}) then function $G_{n+1} = D_n G_n$ is again a solution. This completes the proof of the Lemma. 

Now turn back to the proof of the theorem. Since for arbitrary integer $k$ the set $G^{(1)}_k, G^{(2)}_k, \ldots, G^{(N)}_k$ constitutes a set of independent in principal solutions of the system (\ref{pr3_eq4}) of the order $m$ then any other solution of the order $m$ is a function of these basic solutions. In other words there exists a function $H_n = \left\{ H^{(1)}_n, H^{(2)}_n, \ldots, H^{(N)}_n \right\}$ depending also on $n$ such that
\begin{equation}	\label{pr3_eq9}
G_{n+k} = H_{n+k} (G_k).
\end{equation}
Due to the chain rule we can easily derive a relation for the Jacobians
\begin{equation}	\label{pr3_eq10}
\frac{\partial G_{n+k}}{\partial u_{n,[m]}} = \frac{\partial H_{n+k}}{\partial G_n} \frac{\partial G_n}{\partial u_{n,[m]}}
\end{equation}
that obviously implies that Jacobian $\frac{\partial H_{n+k}}{\partial G_k}$ doesn't vanish. Indeed due to the property of the independence in principal we have that
$\frac{\partial G_{n+k}}{\partial u_{n,[m]}} \neq 0$ and $\frac{\partial G_n}{\partial u_{n,[m]}} \neq 0$. Let us take $k=0$ and find by inversion of the relation (\ref{pr3_eq9}) that $G_0 = H^{-1}_n (G_n)$ and $G_0 =  H^{-1}_{n+1}(G_{n+1}) = D H^{-1}_n (G_n)$. It follows from these two relations that function $I = H^{-1}_n (G_n)$ is an integral of the system (\ref{main_sys}). Indeed we have $DI=I$.

 Theorem 3.3. is proved.

Thus we are ready to formulate the main result of the article.

{\bf Theorem 3.4 (Algebraic criterion of Darboux integrability).}
{\it System (\ref{main_sys}) is integrable in the sense of Darboux if and only if both its characteristic algebras $L_x$ and $L_n$ have finite dimension.}

\section{Construction of a set of independent integrals of minimal orders. Illustrative Examples.}

In this section, we show two examples demonstrating an algorithm for defining the characteristic Lie-Rinehart algebra. We then discuss the subsequent application of algebra to the computation of the set of integrals of minimal orders. The reader can find a large number of examples in \cite{HabibullinKhakimova}.

{\bf Example 4.1.} Let us consider a scalar equation \cite{HZS2010}
\begin{equation} \label{eq4_1}
u_{n+1, x} = u_{n,x} + u^2_{n+1} + u_{n+1} - u^2_n - u_n.
\end{equation}
We apply the method of characteristic algebras to find its $x$-integral. First we write down characteristic operator. In this case it can be presented as
\begin{equation} \label{eq4_2}
K_0 = u_{n,x} \frac{\partial}{\partial u_n} + (u_{n,x} + f_n) \frac{\partial}{\partial u_{n+1}} + (u_{n,x} + g_n) \frac{\partial}{\partial u_{n-1}} + \cdots,
\end{equation}
where $f_n = u^2_{n+1} + u_{n+1} - u^2_n - u_{n}$, $g_n = u^2_{n-1} + u_{n-1} - u^2_n - u_{n}$. According to the scheme above we have to describe the Lie-Rinehart algebra $L_x$ generated by $K_0$ and $X = \frac{\partial}{\partial u_{n,x}}$. We rewrite (\ref{eq4_2}) in a more convenient form $K_0 = u_{n,x} \tilde{X} + K$, where
\begin{equation*}
\tilde{X} = \sum^{+\infty}_{-\infty} \frac{\partial}{\partial u_k}, \quad K = \sum^{+\infty}_{-\infty} (u^2_{n+k} + u_{n+k} - u^2_n - u_n) \frac{\partial}{\partial u_{n+k}},
\end{equation*}
and then find $K_1 = \left[ \tilde{X}, K \right] = \sum^{+\infty}_{-\infty} (u_{n+k} - u_n) \frac{\partial}{\partial u_{n+k}}$. Since $x$-integral does not depend on $u_{n,x}$ we concentrate on the subalgebra $L_0$ generated by the operators $\tilde{X}$ and $K$. Subalgebra is of dimension three, operators $\tilde{X}, K, K_1$ constitute its basis, since we have obviously $[\tilde{X}, K_1] = 0$, $[K_1, K] = 2 K - 2 u_{n,x} \tilde{X} - (2 u_n + 1) K_1$. Finalizing the reasoning above we can conclude that $x$-integral $J$ of the equation (\ref{eq4_1}) should satisfy a system of three equations
\begin{equation} \label{eq4_3}
K_1 J = 0, \quad K J = 0, \quad \tilde{X} J = 0.
\end{equation}
Without loosing generality we can set $J = J(u_n, \ldots, u_{n+m})$. Note that the order $m$ of a nontrivial integral is greater than two. Indeed, for $J(u_n, u_{n+1}, u_{n+2})$ we have from (\ref{eq4_3}) a system of linear equations
\begin{eqnarray*}
J_u + J_{u_1} + J_{u_2} = 0, \\
(u^2_{n+1} + u_{n+1} - u^2_n - u_n) J_{u_1} + (u^2_{n+2} + u_{n+2} - u^2_{n+1} - u_{n+1}) J_{u_2} = 0,\\
(u_{n+1} - u_n) J_{u_1} + (u_{n+2} - u_n) J_{u_2} = 0
\end{eqnarray*}
that implies $J_u = J_{u_1} = J_{u_2} = 0$. Therefore $J$ is a trivial integral depending only on $n$. We can find a nontrivial integral by taking $m = 3$. In this case we get
\begin{equation*}
J = \frac{(u_{n+3} - u_{n+1})(u_{n+2} - u_n)}{(u_{n+3} - u_{n+2})(u_{n+1} - u_n)}
\end{equation*}
which is evidently an integral of the minimal order. For this equation the $n$-integral is easily found, since the equation (\ref{eq4_1}) is rewritten as $u_{n+1, x}- u^2_{n+1} - u_{n+1} = u_{n,x} - u^2_n - u_n$. Therefore $I=u_{n,x} - u^2_n - u_n$ is an $n$-integral of the first order. 

{\bf Example 4.2. } Let us consider the following system
\begin{equation}		\label{eq4_4}
u^0_{n+1,x} = u^0_{n,x} + e^{u^0_{n+1} - u^1_n}, \quad u^1_{n+1,x} = u^1_{n,x} - e^{u^0_{n+1} - u^1_n}.
\end{equation}
Assume that function $H(u^0_n, u^1_n, u^0_{n+1}, u^1_{n+1}, \ldots)$ is an $x$-integral for the system (\ref{eq4_4}). Then by definition we get
$D_x H(u^0_n, u^1_n, u^0_{n+1}, u^1_{n+1}, \ldots) = 0$. Let us apply $D_x$ to $H$, then
\begin{eqnarray*}
\fl D_x H = K_0 H = \left( u^0_{n,x} \frac{\partial}{\partial u^0_n} + u^1_{n,x} \frac{\partial}{\partial u^1_n} + \right. \\
\left. + (u^0_{n,x} + e^{u^0_{n+1} - u^1_n}) \frac{\partial}{\partial u^0_{n+1}} + (u^1_{n,x} - e^{u^0_{n+1} - u^1_n}) \frac{\partial}{\partial u^1_{n+1}} + \cdots \right),
\end{eqnarray*}
In addition we also have two equations:
\begin{equation*}
X_1 H = 0, \quad X_2 H = 0,
\end{equation*}
where $X_1 = \frac{\partial}{\partial u^0_{n,x}}$, $X_2 = \frac{\partial}{\partial u^1_{n,x}}$. Operator $K_0$ is represented as a linear combination of the vector fields:
\begin{equation*}
K_0 = u^0_{n,x} Y_1 + u^1_{n,x} Y_2 + W,
\end{equation*}
where
\begin{eqnarray*}
Y_1 = \frac{\partial}{\partial u^0_n} + \frac{\partial}{\partial u^0_{n+1}} + \frac{\partial}{\partial u^0_{n+2}} + \cdots, \\
Y_2 = \frac{\partial}{\partial u^1_n} + \frac{\partial}{\partial u^1_{n+1}} + \frac{\partial}{\partial u^1_{n+2}} + \cdots,\\
W = e^{u^0_{n+1} - u^1_n} \frac{\partial}{\partial u^0_{n+1}} -  e^{u^0_{n+1} - u^1_n} \frac{\partial}{\partial u^1_{n+1}} + \\
+ \left( e^{u^0_{n+1} - u^1_n} + e^{u^0_{n+2} - u^1_{n+1}} \right) \frac{\partial}{\partial u^0_{n+2}} - \left( e^{u^0_{n+1} - u^1_n} + e^{u^0_{n+2} - u^1_{n+1}} \right) \frac{\partial}{\partial u^1_{n+2}}.
\end{eqnarray*}
Thus we have
\begin{equation}  \label{eq4_5}
Y_1 H = 0, \quad Y_2 H = 0, \quad W H = 0.
\end{equation}
Commutators of operators $Y_1, Y_2, W$ satisfy the following relations:
\begin{equation*}
[Y_1, Y_2] = 0, \quad [Y_1, W] = W, \quad [Y_2, W] = - W.
\end{equation*}
At the first step we are looking for solutions of system (\ref{eq4_5}) of the form $H = H(u^0_n, u^1_n, u^0_{n+1}, u^1_{n+1})$. Then $H$ must satisfy the system:
\begin{equation*}
\frac{\partial H}{\partial u^0_n} + \frac{\partial H}{\partial u^0_{n+1}} = 0, \quad \frac{\partial H}{\partial u^1_n} + \frac{\partial H}{\partial u^1_{n+1}} = 0, \quad \frac{\partial H}{\partial u^0_{n+1}} - \frac{\partial H}{\partial u^1_{n+1}} = 0.
\end{equation*}
The system has a solution of the form
\begin{equation*}
H = P(u^1_{n+1} + u^0_{n+1} - u^0_n - u^1_n).
\end{equation*}
Therefore we can chose the integral
\begin{equation}  \label{eq4_6}
J_1 = u^1_{n+1} + u^0_{n+1} - u^0_n - u^1_n.  
\end{equation}

In order to find two functionally independent solution of system (\ref{eq4_5}) we should consider the function $H$ of the form $\bar{H} = \bar{H}(u^0_n, u^1_n, u^0_{n+1}, u^1_{n+1}, u^0_{n+2})$. To find $H$ we have the following system:
\begin{eqnarray*}
\frac{\partial \bar{H}}{\partial u^0_n} + \frac{\partial \bar{H}}{\partial u^0_{n+1}} + \frac{\partial \bar{H}}{\partial u^0_{n+2}}  = 0, \\
\frac{\partial \bar{H}}{\partial u^1_n} + \frac{\partial \bar{H}}{\partial u^1_{n+1}}  = 0, \\
e^{u^0_{n+1}- u^1_n} \left( \frac{\partial \bar{H}}{\partial u^0_{n+1}} - \frac{\partial \bar{H}}{\partial u^1_{n+1}} \right)+ (e^{u^0_{n+1} - u^1_n} + e^{u^0_{n+2} - u^1_{n+1}}) \frac{\partial \bar{H}}{\partial u^0_{n+2}} = 0.
\end{eqnarray*}
This system has a solution of the form: 
\begin{equation*}
\bar{H} = \bar{H}(u^1_{n+1}-u^1_n + u^0_{n+1} - u^0_n, e^{u^0_{n+1} - u^0_n} + e^{2u^0_{n+1} - u^0_n - u^0_{n+2} + u^1_{n+1} - u^1_n}).
\end{equation*}
We can chose the following integral:
\begin{equation*}
\fl J_2 = e^{-(u^1_{n+1}-u^1_n + u^0_{n+1} - u^0_n)} \left( e^{u^0_{n+1} - u^0_n} + e^{2u^0_{n+1} - u^0_n - u^0_{n+2} + u^1_{n+1} - u^1_n} \right) =
e^{u^1_n - u^1_{n+1}}+e^{u^0_{n+1} - u^0_{n+2}} .
\end{equation*}
Thus we obtain two independent $x$-integrals of the minimal orders $n_1=1$ and $n_2=2$
\begin{equation*}
J_1 = u^1_{n+1} + u^0_{n+1} - u^0_n - u^1_n,\quad \mbox{and}\quad J_2 = e^{u^1_n - u^1_{n+1}}+e^{u^0_{n+1} - u^0_{n+2}} 
\end{equation*}
of system (\ref{eq4_4}).

\section{Conclusion}
It was observed in \cite{HabibullinKhakimova} that every known three-dimensional integrable equation of the form (\ref{eq0}) possesses an infinite set of the Darboux integrable reductions that are systems of differential-difference equations with two independent variables $x$ and $n$. This circumstance is very important from the integrable classification viewpoint, since the Darboux integrable systems are closely related to the Lie-Rinehart algebras. In the present article we proved that a finite system of differential-difference hyperbolic type equations is integrable in the sense of Darboux if and only if both its characteristic Lie-Rinehart algebras are of finite dimension. This fact creates a good basis for the development of an algebraic algorithm for the classification of equations of the form (\ref{eq0}) with three independent variables.  

\section*{Acknowledgements}
The research was supported by a grant from the Russian Science Foundation\\ No.~21-11-00006, https://rscf.ru/project/21-11-00006/.


\section*{References}

\begin{thebibliography}{30}

\bibitem{ZhS2001} Zhiber A V, Sokolov V V 2001 Exactly integrable hyperbolic equations of Liouville type \textit{Russian Math. Surveys} \textbf{56}:1, 61--101.



\bibitem{Goursat1} Goursat E 1899 \textit{ Recherches sur quelques  $\acute{e}$quations aux d$\acute{e}$riv$\acute{e}$es partielles du second ordre}, Annales de la facult$\acute{e}$ des Sciences de l'Universit$\acute{e}$ de Toulouse $2^e$ s$\acute{e}$rie, tome 1, n$^0$ 1,  31--78.

\bibitem{Goursat2}
Goursat E 1899 \textit{ Recherches sur quelques  $\acute{e}$quations aux d$\acute{e}$riv$\acute{e}$es partielles du second ordre}, Annales de la facult$\acute{e}$ des Sciences de l'Universit$\acute{e}$ de Toulouse $2^e$ s$\acute{e}$rie, tome 1, n$^0$ 1 (1899), 79--163.

\bibitem{ves1} Vessiot E 1939   Sur les equations aux derivees partialles du second order,
  $ F(x,y,p,q,r,s,t)=0,$ inteqrables
par la methode de Darboux, \textit{J. Math. Pure Appl.} \textbf{18}, 1--61.

\bibitem{ves2} Vessiot E 1942  Sur les equations aux derivees partialles du second order,
  $ F(x,y,p,q,r,s,t)=0,$ integrables
par la methode de Darboux, \textit{J. Math. Pure Appl.} \textbf{21}, 1--68.

\bibitem{ShabatYam81} Shabat A B, Yamilov R I 1981 Exponential systems of type I and the Cartan matrices (in Russian), Preprint, Bashkirian Branch Acad. Sci. USSR, Ufa 

\bibitem{LSS} Leznov A N, Smirnov V G,  Shabat A B 1982 The group of internal symmetries and the conditions of integrability of two-dimensional dynamical systems \textit{Theor. Math. Phys.} \textbf{51}:1, 322--330.

\bibitem{ZMHS} Zhiber A V, Murtazina R D, Habibullin I T, Shabat A B 2012 Characteristic Lie Rings and Non-linear Integrable Equations (in Russian), Inst. Computer Studies, Moscow 

\bibitem{HPZ2009} Habibullin I, Zheltukhina N, Pekcan A 2009 Complete list of Darboux integrable chains of the form $t_{1,x}=t_x+d(t,t_1)$. \textit{J. Math. Phys.} \textbf{50}, 1--23.




\bibitem{Rinehart} Rinehart G 1963 Differential forms for general commutative algebras \textit{Trans. Amer. Math. Soc.} \textbf{108}, 195--222.

\bibitem{Million} Millionshchikov D 2018 Lie Algebras of Slow Growth and Klein-Gordon PDE \textit{Algebr. Represent. Theor.} \textbf{21}, 1037--1069.
%https://doi.org/10.1007/s10468-018-9794-4.

\bibitem{MillionSmirnov} Millionshikov D V, Smirnov S V 2021 Characteristic algebras and integrable systems of exponential type \textit{Ufa Math. J.} \textbf{13}:2, 44--73.





\bibitem{Anderson} Anderson I M, Kamran N 1997 The variational bicomplex for hyperbolic second-order scalar partial differential equations in the plane \textit{Duke Math. J.} \textbf{87}:2, 265--319.

\bibitem{AdlerStartsev1999} Adler V E, Startsev S Y 1999 Discrete analogues of the Liouville equation \textit{Theor. Math. Phys.} \textbf{121}:2, 1484--1495

\bibitem{Smirnov2015} Smirnov S V 2015 Darboux integrability of discrete two-dimensional Toda lattices \textit{Theor. Math. Phys.} \textbf{182}:2, 189--210.

\bibitem{GarYam19} Garifullin R N, Yamilov R I 2019 On series of Darboux integrable discrete equations on square lattice \textit{Ufa Math. J.} \textbf{11}:3, 100--109.

\bibitem{GarHab21} Garifullin R N, Habibullin I T 2021 Generalized symmetries and integrability conditions for hyperbolic type semi-discrete equations \textit{J. Phys. A: Math. Theor.} \textbf{54}, 205201. 

\bibitem{Demskoy10} Demskoi D K 2010 Integrals of open two-dimensional lattices \textit{Theor. Math. Phys.} \textbf{163}, 466--471.

\bibitem{ZhZh21} Zheltukhin K, Zheltukhina N A 2021 On discretization of Darboux Integrable Systems admitting second-order integrals \textit{Ufa Math. J.} \textbf{13}:2, 176--192.


\bibitem{HabibullinKhakimova} Habibullin I T, Khakimova A R 2021 Characteristic Lie Algebras of Integrable Differential-Difference Equations in 3D \textit{J. Phys. A: Math. Theor.} \textbf{54}:29, 295202
%https://doi.org/10.1088/1751-8121/ac070c

\bibitem{FNR} Ferapontov E V, Novikov V S, Roustemoglou I 2015 On the classification of discrete Hirota-type equations in 3D \textit{Int. Math. Res. Not. IMRN}  \textbf{13} 4933--4974




\bibitem{H2013}  Habibullin I 2013 Characteristic Lie rings, finitely-generated modules and integrability conditions for $(2+ 1)$-dimensional lattices \textit{Physica Scripta} \textbf{87}, 065005 

\bibitem{HP2017} Habibullin I, Poptsova M 2017 Classification of a Subclass of Two-Dimensional Lattices via Characteristic Lie Rings \textit{SIGMA} \textbf{13}, 26 pp.  

\bibitem{HabKuzTMPh} Habibullin I T, Kuznetsova M N 2020 A classification algorithm for integrable two-dimensional lattices via Lie–Rinehart algebras \textit{Theor. Math. Phys.} \textbf{203} 569--581.

\bibitem{HKS2020} Habibullin I T, Kuznetsova M N, Sakieva A U 2020 Integrability conditions for two-dimensional Toda-like equations \textit{J. Phys. A: Math. Theor.} \textbf{53}:39, 395203. 

\bibitem{FHKN2020} Ferapontov E V, Habibullin I T, Kuznetsova M N, Novikov V S 2020 On a class of 2D integrable lattice equations \textit{J. Math. Phys.} \textbf{ 61}:7, 073505.  


\bibitem{KostriginaDiss} Kostrigina O S 2011, Nonlinear hyperbolic systems integrable by Darboux, dissertation. 
	
%\bibitem{1} Allwood JM, Cullen JM. 2011 \textit{Sustainable materials:  with both eyes open}.
%Cambridge, UK: UIT Cambridge. See \href{http://www.withbotheyesopen.com}{http://www.withbotheyesopen.com}.






\bibitem{sigma2005} Habibullin I 2005 Characteristic Algebras of Fully Discrete Hyperbolic Type Equations {\it SIGMA} \textbf{1}, 023.

\bibitem{ZhKuz21} Kuznetsova M N, Zhiber A V 2021 Integrals and characteristic Lie rings of semidiscrete systems of equations \textit{Ufa Math. J.} \textbf{13}:2, 25--35.

\bibitem{HZS2010} Habibullin I,  Zheltukhina N, Sakieva A  2010 On Darboux-integrable semi-discrete chains \textit{J. Phys. A: Math. Theor.} \textbf{43}:43, 434017.

\end{thebibliography}
\end{document}